# Impact of ALD-Deposited Ultrathin Nitride Layers on Carrier Lifetimes and Photoluminescence Efficiency in CdTe/MgCdTe Double Heterostructures

Haris Naeem Abbasi[1,a], Xin Qi[2,a], Zheng Ju[2], Zhenqiang Ma[1,*] and Yong-Hang Zhang[2,*],

[1] Department of Electrical and Computer Engineering, University of Wisconsin-Madison, Madison, WI, 53706, USA

[2] School of Electrical, Computer and Energy Engineering, Arizona State University, Tempe, AZ, 85287, USA

[a] These authors contributed equally.

* Authors to whom correspondence should be addressed: mazq@engr.wisc.edu or yhzhang@asu.edu,

## Abstract

This work evaluates the passivation effectiveness of ultrathin nitride layers ($SiN_x$, AlN, TiN) deposited via atomic layer deposition on CdTe/MgCdTe double heterostructures for solar cell applications. Time-resolved photoluminescence and photoluminescence measurements revealed enhanced carrier lifetimes and reduced surface recombination, indicating improved passivation effectiveness. These results underscore the potential of $SiN_x$ as a promising passivation material to improve the efficiency of CdTe solar cells.

## 1. Introduction

Thin-film solar cells with high power density and specific power are desirable for terrestrial and space applications. Thin-film CdTe solar cells constitute about 5% of the global PV market and hold a record lab efficiency of 22.6% [1-3]. Compared to silicon solar cells with an efficiency of 26.8% efficiency [4, 5], there is still large room for improvement in CeTe cells. Polycrystalline CdSeTe solar cells' open circuit voltage ($V_{OC}$) is limited by the low p-type doping concentration and short minority carrier lifetime [4]. Single crystalline CdTe/MgCdTe double heterostructures (DHs) offer a model system to study the fundamental limits of CdTe solar cells, such as enhanced carrier lifetimes and increased open-circuit voltage ($V_{OC}$) [5-7], and the effectiveness of the thin MgCdTe to suppress surface recombination due to thermionic emission and tunneling processes [8].

Surface passivation has been extensively studied on crystalline silicon (c-Si) [9-11], GaAs [12], copper indium gallium diselenide (CIGS) [13], and CdTe [2, 14-17] based solar cell technologies. The surface passivation can also serve as contact passivation. For effective passivating contacts, films need to be significantly thinner (<< 5 nm) compared to traditional passivation layers that are often several tens of nanometers thick, ensuring improved charge carrier transport [18, 19]. A better understanding of interface charge carrier transport can be achieved by considering film thickness, and distinguishing whether quantum tunneling or transport via pinholes is the dominant mechanism [20, 21].

Metal oxides deposited using atomic layer deposition (ALD) have been used as a passivation layer for crystalline silicon surfaces [10], ultrathin tunneling layers [22] or hydrogenating caps for contact [23]. Recent studies have demonstrated that applying a thin layer of oxide for passivation can significantly increase the $V_{OC}$, leading to a notable improvement in solar cell efficiency [24, 25]. In addition, our recent study on ultra-thin ALD dielectric materials has observed increased carrier lifetime after the thin film deposition [2]. Among those different ALD thin films, ultra-thin TiN (~1.2 nm), which was deposited by an $N_2$ plasma-assisted process, has shown the best performance. Therefore, other nitrides such as $SiN_x$ and AlN deposited by similar processes are worth studying as a passivation layer.

Aside from enhancing the solar cell efficiency by reducing surface recombination, the ultrathin ALD passivation layer can also serve as a tunneling barrier for grafted tandem solar cells, integrating CdTe/MgCdTe DH with materials like Si and Ge [26-28]. This approach, supported by semiconductor nanomembrane grafting techniques, promises higher efficiency in lattice-mismatched heterojunctions [29-32]. In addition, efficient interfacial passivation is crucial to mitigate defects and improve the performance of tandem solar cells, underscoring the importance of studying various ALD dielectric layers for future advancements.

In this study, we investigated how various ultrathin (< 2 nm) ALD nitrides (TiN, $SiN_x$, and AlN) passivate the monocrystalline CdTe/MgCdTe DHs. Time-resolved photoluminescence (TRPL) and photoluminescence (PL) measurements were carried out to examine the optoelectronic properties of the CdTe/MgCdTe DHs. Enhanced carrier lifetime and PL efficiency were observed in DHs with $SiN_x$, TiN, and AlN passivation. The findings present a viable strategy for further mitigating undesired surface recombination in CdTe/MgCdTe DHs, showcasing an effective approach to enhance device performance.

## 2. Experiment

The layer structure of the CdTe/MgCdTe DH is the same as that of the best solar cells [6, 7]. The DH was grown on a 600 µm thick InSb wafer by molecular beam epitaxy (MBE) at 280°C. The CdTe layer acts as an absorber for photons with energies above its 1.5 eV bandgap. The MgCdTe barriers prevent carriers from escaping to the surface, therefore suppressing surface recombination.

This study investigated the passivation effect of various ALD dielectric layers, such as $SiN_x$, AlN, and TiN, on CdTe/MgCdTe DHs. Before dielectric layer deposition, samples underwent a thorough cleaning process using acetone, isopropyl alcohol, and de-ionized water, for 10 minutes each step. The native oxide layer on the sample surface can be removed using de-ionized water, according to the paper by Choi et al. in the 80's [33]. Each nitride layer was then deposited onto a piece of sample with a fresh surface. A schematic diagram of the process flow is shown in **Figure 1**.

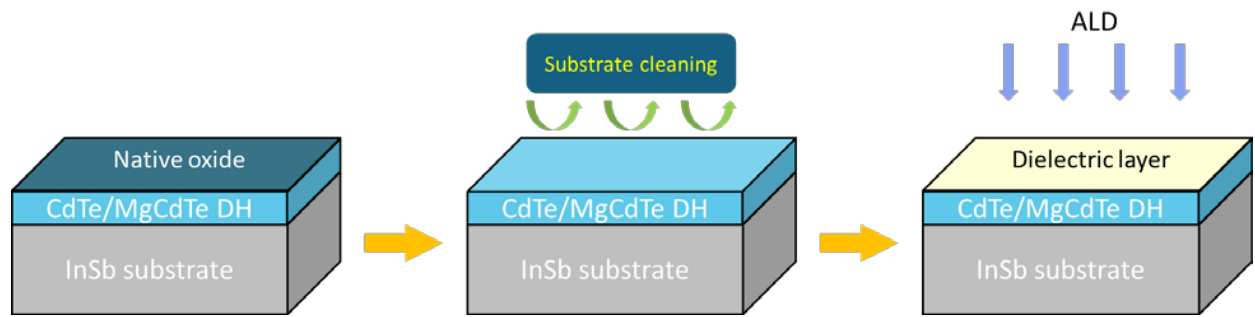

FIG. 1. Schematic diagram of sample cleaning and dielectric layer deposition.

Ultra-thin films of $SiN_x$ and TiN were deposited by ALD using the tris(dimethylamino)silane (TDMASI), and tetrakis(dimethylamido)titanium (TDMAT) precursors, respectively, along with the $N_2$ plasma. In contrast, the AlN film was deposited using the trimethylaluminum (TMA) precursor and $N_2$-$H_2$ co-reactant. To deposit the films of 1.2 nm thickness, 66, 15, and 20 ALD cycles were performed for $SiN_x$, AlN, and TiN, respectively, based on the growth per cycle rates provided by the company ($SiN_x$: ≈0.15Å per cycle, AlN: ≈0.8 Å per cycle, TiN: ≈0.6 Å per cycle). In addition, films with 0.5 nm thickness were also deposited using ALD for comparison. The summary of the ALD parameters is displayed in Table 1.

The design of the ~1.2 nm thick nitride films is driven by the previous passivation study in which ~1.2 nm TiN showed the best performance [2]. In addition, the tunneling regime (0.9 – 2.2 nm) ultrathin layers for the passivation of the crystalline Si surface observed increased carrier lifetimes compared to thicker layers [11]. Therefore, the passivation effect of thinner (~0.5 nm) nitride films was also studied.

Table 1. Summary of ALD parameters in this study.

| ALD dielectric | ALD instrument | Deposition rate per cycle | No of cycles | Precursors Applied | Deposition temperature (°C) | Estimated thickness (nm) |
|---|---|---|---|---|---|---|
| $SiN_x$ | CLAS Fiji | 0.15 Å | 33/66 | BTBAS and N2 Plasma | 300 | 0.5/1.2 |
| AlN | CLAS Fiji | 0.8 Å | 8/15 | TMA and N2-H2 plasma | 250 | 0.5/1.2 |
| TiN | CLAS Fiji | 0.6 Å | 10/20 | TDMAT and N2 plasma | 275 | 0.5/1.2 |

## 3. Results and Discussion

TRPL analysis was conducted to measure carrier lifetimes. A 532 nm pulse laser with a 2 MHz repetition rate was used for sample excitation. The detection wavelength was set at 821 nm, with a resolution of 16 nm. As shown in **Figure 2**, the PL decay times were by fitting the TRPL curves using a single exponential decay model, which indicated the corresponding carrier lifetimes. The effective carrier lifetimes for all the samples are presented in **Table 2**. Enhanced carrier lifetimes were observed in DHs with ~1.2 nm $SiN_x$ and TiN passivation layers, and the carrier lifetime in the sample with $SiN_x$ passivation is around 1.8 times higher than the reference sample. Although the ~1.2 nm AlN passivation decreased the carrier lifetime, an increased carrier lifetime was observed for DH with ~0.5 nm AlN passivation, as shown in **Figure S1**.

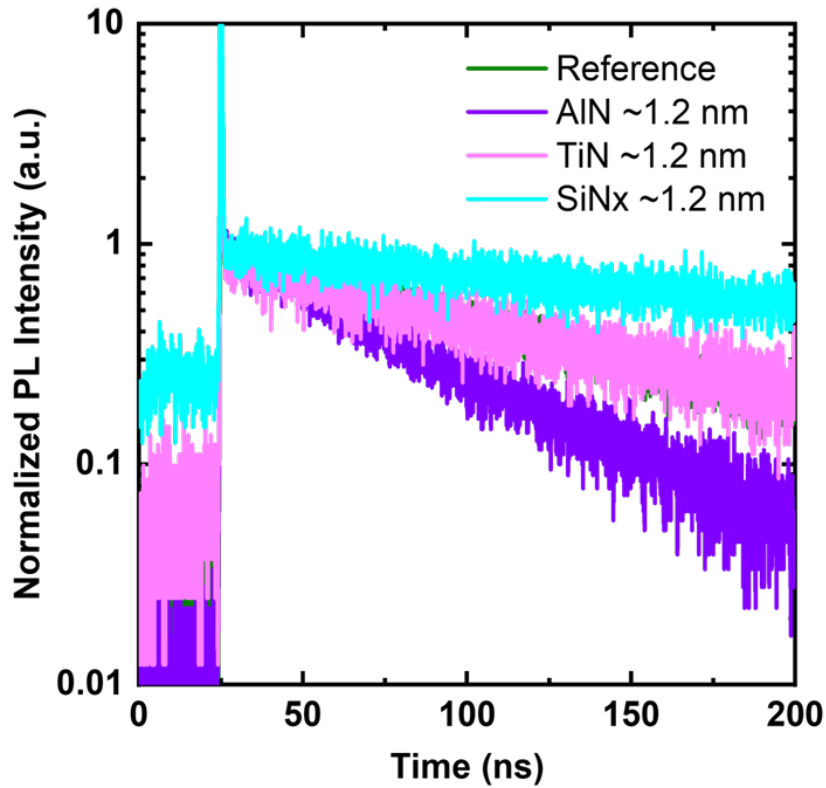

FIG. 2. A set of typical normalized TRPL decay curves of CdTe/MgCdTe DHs with different passivation materials.

Table 2. Summary of PL and TRPL measurements and the corresponding intensity and the effective carrier lifetimes.

| Passivation | Reference (w/o) | AlN | | TiN | | SiN$_x$ | |
|---|---|---|---|---|---|---|---|
| | | 0.5 nm | 1.2 nm | 0.5 nm | 1.2 nm | 0.5 nm | 1.2 nm |
| Carrier lifetime (ns) | 110 | 105 | 61 | 81 | 129 | 208 | 310 |
| PL intensity (a.u.) | 10794.49 | 17215.45 | 11123.19 | 10523.24 | 15043.4 | 24858.62 | 24464.82 |

According to the carrier recombination theory, the measured effective carrier lifetime $\tau_{\text{eff}}$ can be expressed using the following equation [34, 35].

$$\frac{1}{\tau_{\text{eff}}} = \frac{1}{\tau_{\text{surface}}} + \frac{1}{\tau_{\text{bulk}}} \quad (1)$$

where $\tau_{\text{bulk}}$ is the bulk carrier lifetime and $\tau_{\text{surface}}$ is the surface recombination lifetime. By assuming that the passivation layers on the DHs did not significantly alter their crystalline structure, the bulk carrier lifetime ($\tau_{\text{bulk}}$) remained constant and the change of measured effective carrier lifetime ($\tau_{\text{eff}}$) mainly attributed to changed surface carrier lifetime ($\tau_{\text{surface}}$). Thus, the increase of carrier lifetime for SiN$_x$ and TiN passivated samples is attributed to the reduction of surface recombination.

In addition, the effectiveness of passivation is characterized by measuring the surface recombination velocity (SRV), which quantifies the rate at which charge carriers recombine at the surface [11].

$$SRV = \frac{W}{2} \left( \frac{1}{\tau_{\text{eff}}} - \frac{1}{\tau_{\text{bulk}}} \right) \quad (2)$$

where $W$ is the thickness of the CdTe absorber layer. The SRV for the SiN$_x$ passivated sample with an absorber thickness of ~1 µm is calculated to be 150 cm·s$^{-1}$. Furthermore, the passivation quality can also be quantified by the recombination current density ($J_0$) which directly correlates with the surface recombination velocity, providing a measure of how effectively a material can prevent carrier recombination [36].

$$SRV = J_0 \left( \frac{N_b + \Delta n}{q n_i^2} \right) \quad (3)$$

where $N_b$ represents the concentration of dopants in the bulk material, $\Delta n$ denotes the increase in carrier concentration, $q$ is the elementary charge, and $n_i$ refers to the intrinsic carrier density in CdTe.

Typically, the SRV decreases by increasing the thickness of the passivation layer as studied by Cui et al. [37] and Gougam et al. [38]. However, the study done by Pain et al. [11] on c-Si showed that tunneling regime passivation (<2 nm) results in low SRV and recombination current density. Consequently, the sample passivated with $SiN_x$ exhibited lower SRV and $J_0$ compared to other passivation layers, making it suitable for creating tandem solar cells using grafting techniques.

The optical characteristics of samples treated with various passivation materials were assessed using steady-state PL at room temperature. A pump laser operating at a 532 nm wavelength was used to excite the samples for this analysis. **Figure 3** compares the PL peak intensities and carrier lifetimes across DHs using various passivation materials. Generally, DHs with longer carrier lifetimes showed stronger PL intensities, indicating enhanced PL efficiency due to reduced non-radiative recombination. The PL intensities of samples passivated with ~1.2 nm $SiN_x$, TiN, and AlN films are about 2.26, 1.28, and 1.13 times that of the reference sample, respectively, suggesting a reduction in non-radiative recombination on the surfaces and improved external quantum efficiency. ALD-$SiN_x$ passivated samples with ~0.5 nm and ~1.2 nm nitride layers showed almost identical PL intensity. In the case of the TiN passivated samples, the PL intensity has surged for a ~1.2 nm passivated sample compared to a ~0.5 nm counterpart. Although the TiN passivated sample outperformed the samples passivated with ALD oxides such as $SiO_2$, $Al_2O_3$, $HfO_2$, $ZrO_2$, and $TiO_2$ in the previous study [2], it has been discovered that nitride-based passivation layers, specifically $SiN_x$ in this study, exhibit even better performance than ALD-deposited TiN. These ALD layers hold significant promise for enhancing CdTe solar cell performance and facilitating their integration into heterogeneous tandem solar cell configurations.

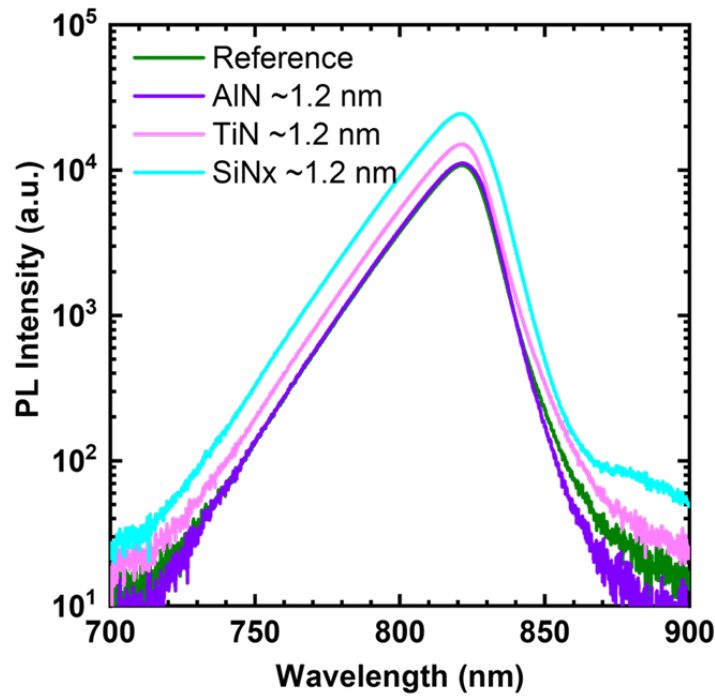

FIG. 3. Room-temperature PL spectra of the studied CdTe/MgCdTe DHs with different passivation materials.

**Figure 4** presents a comparative analysis of the PL peak intensities and carrier lifetimes of double heterostructures utilizing various passivation materials. In general, DHs exhibiting longer carrier lifetimes tend to display enhanced PL intensity, affirming that the reduction of non-radiative recombination significantly contributes to the heightened PL intensity. These findings indicate the effective passivation of the ultrathin nitride layers. In an ideal scenario, a solar cell at an open circuit would emit one photon for each absorbed photon, without any additional nonradiative recombination or photon loss reducing efficiency. The external luminescence efficiency ($\eta_{ext}$) thus serves as an indicator for potential loss mechanisms [39]. **Equation (4)** describes the relationship between the external luminescence efficiency and the implied open circuit voltage ($iV_{OC}$) [40].

$$iV_{OC} = V_{OC-ideal} - \frac{k_B T}{q} |\ln \eta_{ext}|, \qquad (4)$$

in which $V_{OC\text{-ideal}}$ is the maximal voltage that an absorber can generate under the detailed-balance condition. Without these losses, 100% external luminescence and the highest $iV_{OC}$ could be achieved. CdTe/MgCdTe DH with ultra-thin $SiN_x$, AlN, or TiN nitride layers showed increased carrier lifetimes and PL intensity, indicating higher $\eta_{ext}$ and as a result improved $iV_{OC}$. Consequently, solar cells with these passivation layers could achieve higher $V_{OC}$.

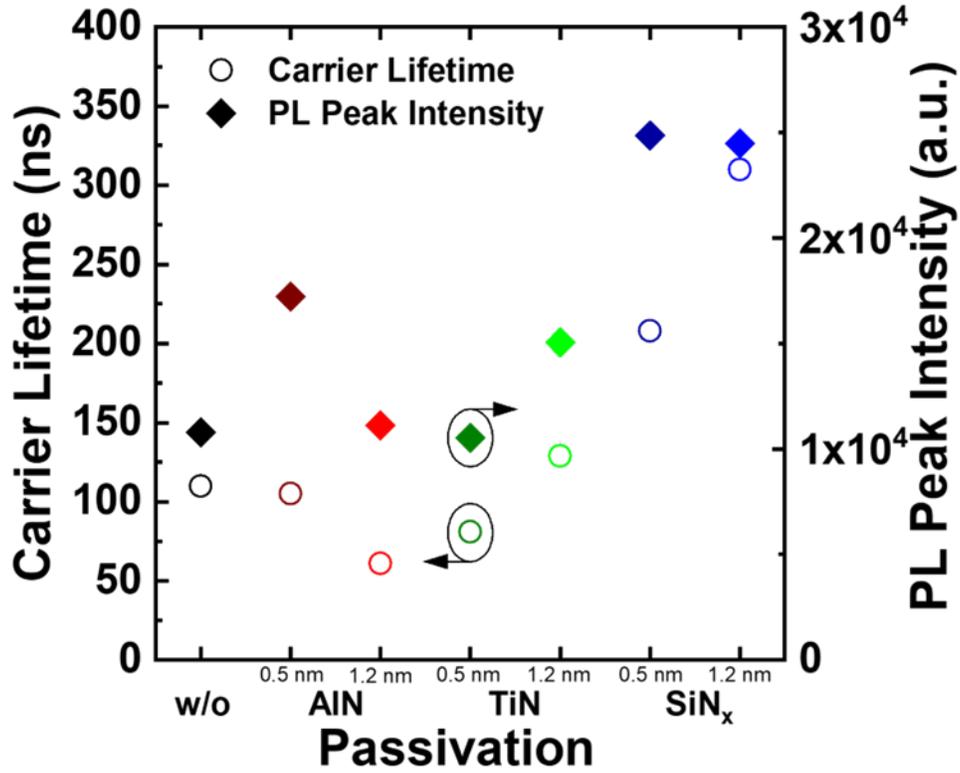

FIG. 4. Combined plots of PL peak intensity and carrier lifetime in the studied DH samples with different passivation materials.

This study showcases the high carrier lifetimes for CdTe/MgCdTe DH samples with $SiN_x$ passivation. Effective passivation in CdTe/MgCdTe DHs can be enhanced by the presence of positive charges in the nitride layer, which helps create an inversion layer on the surface of the DHs [41, 42], contributing to high carrier lifetimes. However, further investigation is needed to identify the nitride passivation effect.

## Conclusion

This study investigates the passivation impact of ultrathin (<2 nm) nitride layers (SiN$_x$, AlN, TiN) deposited via ALD on the CdTe/MgCdTe DHs. Enhanced carrier lifetimes and photoluminescence were observed, indicating a reduction in surface recombination and the potential for higher efficiency of CdTe solar cells. These findings highlight the effectiveness of SiN$_x$ as a superior passivation layer, suggesting its promising role in advancing solar cell technology and enabling the development of high-performance tandem solar cells through semiconductor nanomembrane grafting techniques.

## Acknowledgment

The authors of the Arizona State University acknowledge the grant support from Air Force Research Laboratory (AFRL) (Grant No. FA9453-20-2-0011) and ASU Ultrafast Laser Facility for the TRPL measurement. The authors of the University of Wisconsin–Madison acknowledge the grant support from Air Force Office of Scientific Research (AFOSR) (Grant No. FA9550-21-1-0081) and National Science Foundation (ECCS 2235377).

## Supplementary material

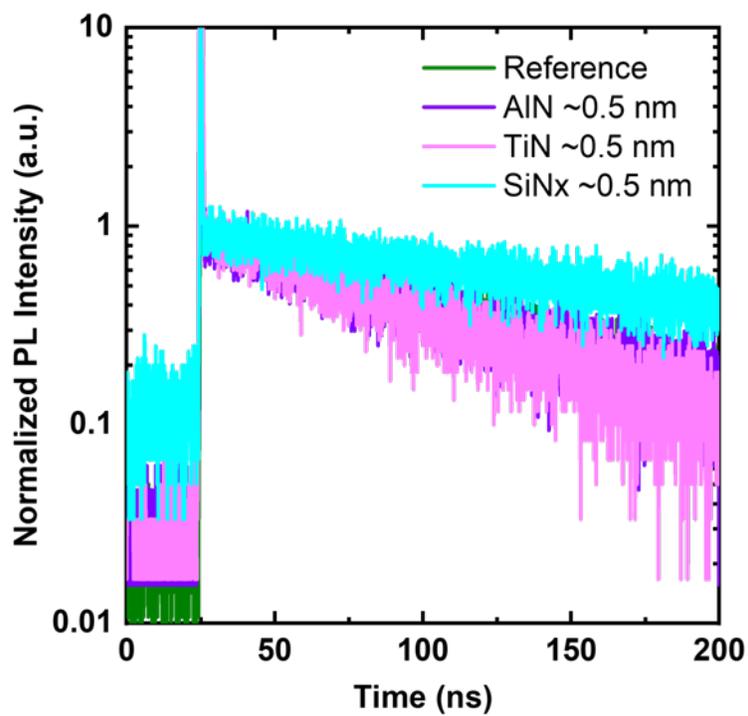

FIG. S1. A set of typical normalized TRPL decay curves of CdTe/MgCdTe DHs with different passivation materials.

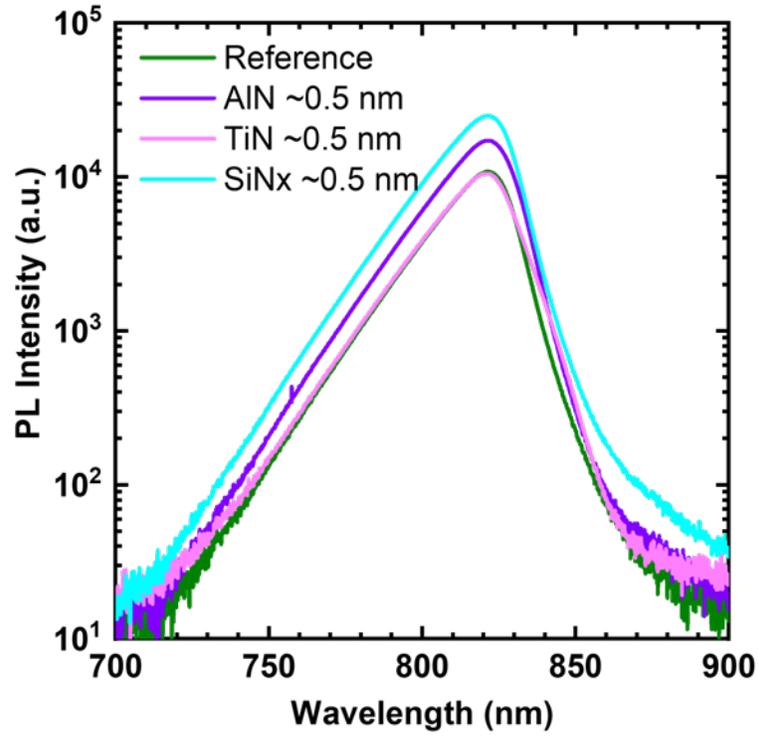

FIG. S2. Room-temperature PL spectra of the studied CdTe/MgCdTe DHs with different passivation materials.